\documentclass[prd,preprintnumbers,nofootinbib,preprint]{revtex4} 
\usepackage{epsfig,graphicx}
\usepackage{dcolumn}
\usepackage{bm}
\usepackage{latexsym}
\usepackage{amsfonts}
\usepackage{amssymb}
\usepackage{amsmath}

\begin{document}

\preprint{IPMU14-0132}

\title{Large Hierarchy from Non-minimal Coupling}

\author{Chunshan Lin}
\email{chunshan.lin@ipmu.jp}
\affiliation{
Kavli Institute for the Physics and Mathematics of the Universe (WPI), Todai Institutes for Advanced Study, University of Tokyo, 5-1-5 Kashiwanoha, Kashiwa, Chiba 277-8583, Japan}

\date{\today}

\begin{abstract}
 In this paper, a model is proposed to solve the gauge hierarchy problem. Beyond the standard model, we introduce an extra scalar field that non-minimally couples to gravity. The fundamental scale is set at weak scale  and Planck scale emerges dynamically by a spontaneous symmetry breaking mechanism. 
\end{abstract}

\maketitle

Our universe is full of mysteries. One of biggest mysteries is the giant hierarchies among Planck scale, weak scale and cosmological constant scale. 

The first type of hierarchy is the so called gauge hierarchy problem, which requires physics beyond the standard model at TeV weak scale. The weak scale is given by the VEV of higgs, which is not natrually stable against radiative correction. If the standard model is valid up to Planck scale, then the minima of higgs potential will be driven by radiative correction to Planck scale. 

One of possible way to solve the gauge hierarchy problem is to introduce supersymmetry. Supersymmetry removes the power law divergence of radiative correction and solve the gauge hierarchy problem, as long as the supersymmetric particles are light enough to satisfy the Barbieri-Giudice criterion\cite{BGcriterion}. However, it still leaves the $\mu$ problem as an open question: why the supersymmetric Higgs mass parameter $\mu$ is so much smaller than the Planck scale?  On the other hand, so far we haven't found any evidence of supersymmetry yet.

Another attempt is to consider the higher-dimensions, in which our universe are confined as a brane\cite{ArkaniHamed:1998rs}\cite{Antoniadis:1998ig}\cite{Gogberashvili:1998vx}\cite{Randall:1999ee}. For example, in the Randall-Sundrum scenario, the large gauge hierarchy is generated by the exponential warping geometry along the extra-dimension\cite{Randall:1999ee}.  However, until now, there isn't any observational evidence of extra-dimension is reported. 

The second type of hierarchy problem is even more profound, which is the so called cosmological constant problem \cite{cc}: why is the energy density of empty space 120 orders of magnitude smaller than the energy density of Planck scale? Even if we only compare to the particle physics scale, the hierarchy is still very large, at least 60 orders of magnitude smaller than several known contributions to it from the Standard Model. Several ideas have been proposed in attempt to solve the cosmological constant problem, see Weinberg's classification in his famous review \cite{cc} (for a newer classification, see \cite{DEbyli}).

In this paper, we aim at gauge hierarchy problem and propose a possible solution in the level of effective theory. As we learn from the textbook, the Fermi's coupling constant in Fermi's  weak interaction theory is suppressed by the square of mass of W boson.  In standard model, the mass of intermediate W boson is generated by a spontaneous symmetry breaking mechanism, which causes the Higgs scalar to have non-trivial vacuum expectation value (VEV).  Inspired by this, we propose a new model that setting up our fundamental scale at weak scale instead of Planck scale. The Planck mass emerges dynamically by the similar spontaneous symmetry breaking mechanism. Beyond the strandard model, we introduce a dilaton type of scalar field non-minimally couples to gravity. The weakness of gravity, or in other word, the largeness of Planck mass is generated by the non-trivial VEV of dilaton. Thus the dilaton symmetry is spontaneously broken in this nontrivial vacuum.
Our action could be written down as follows,
\begin{eqnarray}\label{action}
\int d^{4}x\sqrt{-g}&&\left[\frac{M^{2}}{2}e^{2\phi/\eta}\mathcal{R}-\frac{1}{2}g^{\mu\nu}\partial_{\mu}\phi\partial_{\nu}\phi-\lambda_{h}\left(\phi^{2}-\upsilon_{h}^{2}\right)^{2}\right.\nonumber\\
&&~~~~~~~~\left.-\frac{1}{2}g^{\mu\nu}D_{\mu}H^{\dagger}D_{\nu}H-\lambda_{v}\left(\mid H\mid^{2}-\upsilon_{v}^{2}\right)^{2}-\Lambda\right]~,
\end{eqnarray}
where $M$ is the fundamental scale in $3+1$ dimensional space-time,
$\phi$ is the hidden quasi-dilaton scalar that beyond the standard model, $H$ is the Higgs scalar in standard model, and $\Lambda$
is a bare cosmological constant.  This theory is closely related to the induced gravity \cite{sakharov}\cite{fuji1974}\cite{azee}, wherein one starts with a dynamical spin-2 field coupled to matter sector, and the gravitational coupling constant is dynamically determinded by matter sector. At low energy scale, the space time diffeomorphism invariance remains unbroken, and thus the graviton is still massless in our theory. 

To solve the gauge hierarchy problem, we suggest that funamental scale $M$ should be set at weak scale.  Thus a natural choice for parameters  $\eta$, $\upsilon_h$ and $\upsilon_v$ is that $\eta,\upsilon_h,\upsilon_v\sim\mathcal{O}\left(M\right)$, and dimensionless parameters $\lambda_h,\lambda_v\lesssim1$. 
On the other hand, we leave the bare cosmological constant as a general parameter for the time being. 

Suppose that our early universe started from a quantum era, with scalar field dispersion $\langle\phi^2\rangle>\upsilon_h^2$. In this epoch, scalar field couldn't feel the lump in the potential, and the hierarchy was small at that time. As the universe expands, the scalar dispersion goes down, and scalar field starts to roll down to one of its non-trivial vacua. We assume  that our universe lives in the positive vacuum $\phi\simeq+\upsilon_h$, in which the large hierarchy was generated.  We properly tune the parameters $\upsilon_h/\eta\simeq35$ such that 
\begin{eqnarray}
M_{pl}^{eff}=Me^{\upsilon/\eta}\simeq M\cdot 10^{15},
\end{eqnarray}
and thus the induced gravitational coupling constant is consistent with our observations nowaday. 

In the standard model, the radiative correction drives Higgs mass up to Planck scale, if we assume that standard model is valid up to Planck scale.  In contrary to what happening in standard model, when the energy scale reaches the weak scale in our theory, the symmetry restores since the scalar field $\phi$ receives a temperature induced mass $T^2\phi^2$. The VEV of scalar field  vanishes $\langle\phi\rangle\simeq0$ and the Planck scale is lowered down to the weak scale. Since the Planck scale and weak scale are degenerate, which implies the strong coupling between gravity sector and matter sector. Quantum gravity effect also becomes important and our effective field theory breaks down above the weak scale.

Thus we argue that the radiative correction to Higgs mass should be cut off at the weak scale in our model. It is worth to notice that the way we solve the gauge hierarchy problem is a bit similar to the  Randall-Sundrum scenario \cite{Randall:1999ee}, wherein the fundamental scale on our brane world is set at weak scale due to the warp geometry. 

It is straightforward to derive the  Einstein equations from our action eq.(\ref{action}). Noted that 
\begin{eqnarray}
\delta\mathcal{R}=\delta g^{\mu\nu}\mathcal{R}_{\mu\nu}+g^{\mu\nu}\delta\mathcal{R}_{\mu\nu}.
\end{eqnarray}
In general relativity (GR), the second term on the right hand side of the above equation yields to a total divergence and thus doesn't affect the equation of motion. Here, however, this term must be kept due to the non-minimal coupling between scalar field $\phi$ and gravity. After taking this term into account, the modified Einstein equations read,
\begin{eqnarray}
M^2e^{2\phi/\eta}\left(\mathcal{R}^{\mu\nu}-\frac{1}{2}g^{\mu\nu}\mathcal{R}\right)+M^2\left[\left(e^{2\phi/\eta}\right)^{;\lambda}_{~~;\lambda}g^{\mu\nu}-\left(e^{2\phi/\eta}\right)^{;\mu;\nu}\right]+\Lambda g^{\mu\nu}=T^{\mu\nu}_m+T^{\mu\nu}_{\phi},
\end{eqnarray}
where $;$ denotes the covariant deriative and $T^{\mu\nu}_m$ is the energy momentum tensor of our matter sector, which includes the standard model matter and dark matter. $T^{\mu\nu}_\phi$ is the energy momentum tensor for scalar field $\phi$, 
\begin{eqnarray}
T^{\mu\nu}_\phi=\partial^{\mu}\phi\partial^{\nu}\phi-g^{\mu\nu}\left[\frac{1}{2}g^{\alpha\beta}\partial_{\alpha}\phi\partial_{\beta}\phi+V\left(\phi\right)\right].
\end{eqnarray}

By taking the variation of the action with respect to scalar $\phi$, the equation of motion reads
\begin{eqnarray}\label{eom}
g^{\mu\nu}\nabla_{\mu}\nabla_{\nu}\phi-\frac{\partial V\left(\phi\right)}{\partial \phi}+\frac{M^2}{\eta}e^{2\phi/\eta}\mathcal{R}=0,
\end{eqnarray}
provided that there is no direct coupling between scalar field $\phi$ and standard model fields.  If the scalar field $\phi$ directly couples to standard model fields, then the right hand side of this equation shuld be replaced by $\delta\mathcal{L}_{sm}/\delta\phi-\partial_{\mu}\left(\delta\mathcal{L}_{sm}/\delta\partial_{\mu}\phi\right)$.

As a self-consistency check, let's see if the hidden quasi-dilaton scalar is indeed  trapped and stablized around the vacuum $\phi\simeq\upsilon_h$ during late time cosmic era. According to the equation of motion, at the local minimum we have 
\begin{eqnarray}\label{minima}
4\lambda_{h}\left(\phi^{2}-\upsilon_{h}^{2}\right)\phi-\frac{M^{2}}{\eta}e^{2\phi/\eta}\mathcal{R}=0~.
\end{eqnarray}
In the above equation, we have assumed that during late time epoch the cosmological constant is dominant and $\mathcal{R}$ takes a constant value and thus the scalar field $\phi$ has already settled down at somewhere
\begin{eqnarray}
\phi=\upsilon_h+\delta~,
\end{eqnarray}
 where $\delta$ is a small deviation from the vacuum $\phi=\upsilon_h$. Up to the leading order, the eq.(\ref{minima}) could be approximately rewritten as 
\begin{eqnarray}
8\lambda_{h}\upsilon_{h}^{2}\delta\simeq\frac{M^{2}}{\eta}e^{2\upsilon_{h}/\eta}\mathcal{R}.
\end{eqnarray}
During late time epoch,  Einstein equations tell us 
\begin{eqnarray}
M^{2}e^{2\upsilon_{h}/\eta}\mathcal{R}\simeq4\Lambda,
\end{eqnarray}
thus we have
\begin{eqnarray}
\delta\simeq\frac{\Lambda}{2\lambda_{h}\eta\upsilon_{h}^{2}}.
\end{eqnarray}
As long as the bare cosmological constant satisfies the condition, 
\begin{eqnarray}
\Lambda\ll\lambda_{h}\eta\upsilon_{h}^{3},
\end{eqnarray}
the deviation from the vacuum $\phi\simeq\upsilon_h$ is always very small, i.e. $\delta\ll\upsilon_h$. Even if we include the direct coupling between scalar $\phi$  and standard model fields, $\delta\ll\upsilon_h$ still holds as long as the coupling is weak enough.

The non-minimal coupling between scalar field $\phi$ and gravity generally leads to the violation of equivalence principle. However, at low energy scale, the scalar field $\phi$ is well-trapped at the bottom of potential and thus it is very hard to distinguish between our model and GR. The distinguishable consequences are more likely to appear around or above weak scale in collider.  The mass scale of the new particle associated with this hidden scalar field $\phi$ is also about $TeV$ scale, provided that $35\eta\simeq\upsilon_h\sim\mathcal{O}(M)$, and $\lambda\lesssim1$. Such scale is detectable even at LHC.

Another channel to find the observational signal is to follow the universe backwards in time.  If we trace back the cosmic expansion history,  we would expect a non-negiligible deviation from the vacuum $\phi=\upsilon_h$, due to the higher energy scale at the early time. The shifting of vacuum leads to the running of Newtonian constant, which could be roughly evaluated in terms of the variation per Hubble time,
\begin{eqnarray}
\epsilon_g\equiv\frac{\dot{G}}{HG}= -\frac{\dot{M}_{pl}}{HM_{pl}}=-\frac{2\dot{\phi}}{H\eta}\simeq 2\frac{\Delta\phi}{\eta},
\end{eqnarray}
where $\Delta\phi$ is the scalar field's excursion in one Hubble time. This excursion arises from two effects, one is  the time dependence of Ricci scalar term in the eq.(\ref{eom}), the other is the temperature induced mass term $T^2\phi^2$.  According to the eq. (\ref{minima}), the former could be estimated as 
\begin{eqnarray}
\epsilon_g\sim\frac{\Delta\phi}{\eta}\sim\frac{M^2e^{2\phi/\eta}\mathcal{R}}{\lambda\upsilon_h^2\eta^2}\sim\frac{\rho_{m}}{M^4},
\end{eqnarray}
where $\rho_m$ is the energy density of cold pressureless matter (assuming that cosmological constant term is negligible at that time).
During matter dominant epoch, this variation in $G$ is completely negligible since the $\rho_m\ll M^4$. On the other hand, this variation even vanishes  during the radiation dominant epoch since the Ricci scalar becomes zero. The second source of scalar field's excursion is the temperature induced mass $T^2\phi^2$. The local minima of scalar potential satisfies 
\begin{eqnarray}
4\lambda_{h}\left(\phi_{min}^{2}-\upsilon_{h}^{2}\right)\phi_{min}+cT^2\phi_{min}=0.
\end{eqnarray}
Up to a good approximation, let's equate the excursion of scalar field with the shifting of local minima $\phi_{min}$,
\begin{eqnarray}
\phi_{min}^2=\upsilon_h^2-\frac{cT^2}{4\lambda_h}.
\end{eqnarray}
In the case with low temperature $T\ll \upsilon_h\sim M$, we have  
\begin{eqnarray}
\epsilon_g\sim\frac{\Delta\phi_{min}}{\eta}\sim\frac{cT\Delta T}{\lambda\upsilon_h\eta}\sim\frac{T\Delta T}{M^2},
\end{eqnarray}
where $\Delta T$ is the temperature variation during one Hubble time at radiation dominant epoch, and it takes negative value since temperature goes down as universe expands. We can see that such variation in $G$ could be non-negligible when $T$ is not too small comparing to the weak scale. 

In the case with high temperture $\frac{cT^2}{4\lambda_h}>\upsilon_h^2$, there is only one minimum in the potential, which $\phi_{min}=0$. As we discussed before, in this case our system strongly coupled and effective field theory breaks down.

To have a better understanding in our theory, let's transform our action into Einstein frame, by doing such Weyl rescaling
\begin{eqnarray}
g_{\mu\nu}=\tilde{g}_{\mu\nu}e^{-2\phi/\eta}\mid_{\phi=\upsilon_h+\tilde{\phi}}. 
\end{eqnarray}
Up to leading order, the action could be rewritten as 
\begin{eqnarray}\label{action2}
\int d^{4}x\sqrt{-\tilde{g}}&&\left[\frac{M^{2}}{2}\tilde{\mathcal{R}}\left(\tilde{g}\right)-\frac{1}{2}\tilde{g}^{\mu\nu}\partial_{\mu}\tilde{\phi}^{c}\partial_{\nu}\tilde{\phi}^{c}-4\lambda\upsilon_h^2e^{-2\upsilon_h/\eta}\tilde{\phi}^{c2}-...\right.\nonumber\\
&&~~~~~~\left.-\frac{1}{2}\tilde{g}^{\mu\nu}D_{\mu}H^{c\dagger}D_{\nu}H^{c}-\lambda_{v}\left(\mid H^{c}\mid^{2}-e^{-2\upsilon_{h}/\eta}\upsilon_{v}^{2}\right)^{2}-\Lambda e^{-4\upsilon_{k}/\eta}\right],
\end{eqnarray}
where $\phi^c$ and $H^c$ are canonical normalized Higgs scalars, 
\begin{eqnarray}
\tilde{\phi}^c\equiv\tilde{\phi} e^{-\upsilon_{h}/\eta},\nonumber\\
H^c\equiv H e^{-\upsilon_{h}/\eta},
\end{eqnarray}
and dots stand for the higher order terms in $\tilde{\phi}^c$, which is irrelevant at the low energy scale.  Please notice that the new symmetry breaking scale now is set by 
\begin{eqnarray}
\upsilon_v\to\upsilon_ve^{-\upsilon_{h}/\eta},
\end{eqnarray}
and TeV scale is generated if we choose parameters $\upsilon_{h}/\eta\simeq35$. This result is completely general, any mass parameter in the standard model receives such exponentially suppression,
\begin{eqnarray}
m\to me^{-\upsilon_{h}/\eta}.
\end{eqnarray}

On the other hand, one could read from the action (\ref{action2}) that bare cosmological constant also receives a large exponential suppression factor,
\begin{eqnarray}
\Lambda_{eff}=\Lambda e^{-4\upsilon_{k}/\eta}\simeq \Lambda \cdot10^{-60}.
\end{eqnarray}
So the cosmological constant problem is more or less alleviated in our model. Of course we may still need another $60$ orders of magnitude to solve it. 
Nevertheless, noted that if the bare cosmological constant is set at $\Lambda\sim M^4\cdot10^{-60}\sim\left(TeV\right)^4$, then the effective cosmological constant will be the same order as the one we observe nowaday, and accelerates our cosmic expansion.  Indubitability, it is extremally interesting to ask what kind of physics could play the role of cosmological constant at such scale.

Let's also mention several previous related works. In ref. \cite{cognola},  a dark energy cosmology model was proposed in the modified Gauss-Bonnet gravity, the connection between gauge hierarchy problem and Gauss-Bonnet gravity was also discussed in their paper.  See ref. \cite{Tkach:2012an} for another example in the  same framework of higher-order modifed gravity. See also ref. \cite{Calmet:2007mz} for another possible solution in the framework of Jordan-Brans-Dicke theory. In our case, at low energy scale our theory is just usual standard model and GR. The violation of equivalence principle and some other quantum gravity effects appear around TeV scale.

{\bf Acknowledgments}
The author would like to thank the hospitality of IAS Hong Kong University of Science and Technology, this paper is finalised during author's visiting.  The author would like to thank Y. Cai,  X. Gao, Q. Huang, R. Huo, S. Matsumoto, S. Mukohyama, M. Sasaki,  G. Shiu, T. Suyama, H. Tye,  W. Xue for the useful discussion, and thank Japanese Sake as well, since the idea of this paper came up by itself when the author was drunk.
  This work is supported by the World Premier International Research Center Initiative (WPI Initiative), MEXT, Japan.


\begin{thebibliography}{99}
\bibitem{BGcriterion}
R. Barbieri, G. F. Giudice, Nuclear Physics B306 (1988) 63-76.

\bibitem{ArkaniHamed:1998rs} 
  N.~Arkani-Hamed, S.~Dimopoulos and G.~R.~Dvali,
  Phys.\ Lett.\ B {\bf 429}, 263 (1998)
  [hep-ph/9803315].

\bibitem{Antoniadis:1998ig} 
  I.~Antoniadis, N.~Arkani-Hamed, S.~Dimopoulos and G.~R.~Dvali,
  Phys.\ Lett.\ B {\bf 436}, 257 (1998)
  [hep-ph/9804398].
\bibitem{Gogberashvili:1998vx} 
  M.~Gogberashvili,
  Int.\ J.\ Mod.\ Phys.\ D {\bf 11}, 1635 (2002)
  [hep-ph/9812296].

\bibitem{Randall:1999ee} 
  L.~Randall and R.~Sundrum,
  Phys.\ Rev.\ Lett.\  {\bf 83}, 3370 (1999)
  [hep-ph/9905221].
\bibitem{cc}
S. Weinberg, Reviews of Modern Physics, Vol. 61, No. 1, January 1989.

\bibitem{DEbyli}
M,~Li, X-D,~Li, S, Wang, Y, Wang, Commun.Theor.Phys.56:525-604,2011.

\bibitem{sakharov}
A. D. Sakharov, Sov. Phys. Dokl. {\bf 12}, 1040 (1968).

\bibitem{fuji1974}
Y. Fujii, Phys. Rev. D {\bf 9}, 874 (1974).

\bibitem{azee}
A. Zee, Phys. Rev. Lett. {\bf 42}, 417 (1979).



\bibitem{cognola}
G. Cognola, E. Elizalde, S. Nojiri, S. D. Odintsov, and S. Zerbini, Phys. Rev. D {\bf73}, 084007 (2006).


\bibitem{Tkach:2012an} 
  V.~I.~Tkach,
  Mod.\ Phys.\ Lett.\ A {\bf 27}, 125031 (2012)
  [arXiv:1204.6523 [hep-th]].
\bibitem{Calmet:2007mz} 
  X.~Calmet,
  Phys.\ Rev.\ D {\bf 77}, 047502 (2008)
  [arXiv:0708.2767 [hep-ph]].



\end{thebibliography}
\end{document}